\documentclass[preprint,12pt]{elsarticle}
\journal{Acta Materialia}

\usepackage{graphics,epsfig}
\usepackage{graphicx}
\usepackage{float}
\usepackage{amssymb,amstext,amsmath}
\usepackage{mathtools}
\usepackage{booktabs}
\usepackage{natbib}
\usepackage[latin1]{inputenc}
\usepackage{epstopdf}

\begin{document}

\begin{frontmatter}
\title{Thermodynamic dislocation theory of adiabatic shear banding in steel} 
\author{K.C. Le\,$^a$\footnote{Corresponding author. E-mail: chau.le@rub.de}, T.M. Tran\,$^a$, J.S. Langer\,$^b$}
\address{$^a$\,Lehrstuhl f\"ur Mechanik - Materialtheorie, Ruhr-Universit\"at Bochum, D-44780 Bochum, Germany\\
$^b$\,Department of Physics, University of California, Santa Barbara, California 93106-9530, USA}

\begin{abstract}
The statistical-thermodynamic dislocation theory developed in our earlier studies is used here in an analysis of the experimental observations of adiabatic shear banding in steel by Marchand and Duffy (1988). Employing a small set of physics-based parameters, which we expect to be approximately independent of strain rate and temperature, we are able to  explain experimental stress-strain curves at six different temperatures and four different strain rates.  We make a simple model of a weak notch-like disturbance that, when driven hard enough, triggers shear banding instabilities that are quantitatively comparable with those seen in the experiments.  
\end{abstract}

\end{frontmatter}

\section{Introduction}
\label{Intro} 
Our purpose here is to use the statistical-thermodynamic dislocation theory developed in \cite{LBL-10}-\cite{JSL-arXiv17} to analyze the classic observations of adiabatic shear banding (ASB) by Marchand and Duffy (MD) \cite{Marchand1988}.  The latter authors made stress-strain measurements over a range of substantially different temperatures and shear rates using thin steel tubes bonded to torsional Kolsky bars.  They observed ASB formation at high shear rates and low temperatures. Specifically, they observed abrupt stress drops, large increases of temperature in emerging narrow bands, and strong strain localization leading to crack formation and failure. Our challenge is to predict those behaviors quantitatively using a realistic physics-based theory, and  in this way to obtain additional information about the properties of structural materials.   

Our new ability to interpret data of the kind published in MD \cite{Marchand1988} is related to the fact that, in its latest versions, the statistical thermodynamic dislocation theory is able to predict nonequilibrium behaviors that previously had been beyond the reach of conventional phenomenological methods in this field.  These behaviors include strain hardening \cite{LBL-10,JSL-15}, steady-state stresses over exceedingly wide ranges of strain rates \cite{LBL-10},  Hall-Petch effects \cite{JSL-17a}, thermal softening during deformation \cite{Le2017}, yielding transitions between elastic and plastic responses  \cite{JSL-16,JSL-17a}, and -- most importantly for present purposes -- the competition between thermal and mechanical effects that produces shear banding instabilities \cite{JSL-17,JSL-arXiv17}.

By definition, ``adiabatic'' shear banding is a thermal effect. It happens when the heat generated at a hot spot is unable to flow away from that spot as fast as new heat is being generated there, thus initiating a runaway instability.   These thermal effects were missing in the early versions of the thermodynamic dislocation theory, which were based on data for copper as shown, for example, in Kocks and Mecking \cite{KOCKS-MECKING-03} or in Meyers {\it et al.} \cite{MEYERSetal-95}. There, the thermal conductivity is large enough that heat generation can be neglected and no appreciable thermal softening occurs. However, typical stress-strain curves such as the ones shown for aluminum and steel in \cite{SHIetal-97,ABBODetal-07} and discussed by us in \cite{Le2017} exhibit thermal softening at large strain rates even without undergoing ASB.  When ASB does occur, temperatures within the bands may increase by hundreds of degrees or more, and the thermal forces become one of the principal driving mechanisms.   In order to demonstrate the possibility of ASB formation theoretically, one of us \cite{JSL-16,JSL-17} studied an artificial model with all the same mechanical parameters as copper but with substantially reduced thermal conductivity and an enhanced thermal conversion coefficient.  We will have to be more realistic than this in our analyses of the MD data.  Specifically, we will need to deduce from the MD data a longer list of both mechanical and thermal parameters for their steel alloy than was needed for copper.  For this purpose, we will use the least-squares technique that we described in  \cite{Le2017}.

The thermodynamic dislocation theory is based on two unconventional ideas.  The first of these is that, under nonequilibrium conditions, the atomically slow configurational degrees of freedom of deforming solids are characterized by an effective disorder temperature that differs from the ordinary kinetic-vibrational temperature.  Both of these temperatures are thermodynamically well defined variables whose equations of motion determine the irreversible behaviors of these systems.  The second  principal idea is that entanglement of dislocations is the overwhelmingly dominant cause of resistance to deformation in polycrystalline materials.  It is these two ideas that have led to the successfully predictive theories mentioned above.   

We start in Sec.~\ref{EOM} with a brief summary of the equations of motion to be used here.  Our focus is on the physical significance of the various parameters that occur in them.  We discuss which of these parameters are expected to be material-specific constants, independent of temperature and strain rate, and thus to be key physical ingredients of the theory.  Marchand and Duffy \cite{Marchand1988} provide twelve different stress-strain curves, for six temperatures and four strain rates, for the steel alloy HY-100.  As will be seen, this is enough data for us to use in constructing theories, but these data sets are not immune to experimental uncertainties. The uncertainties  are especially important during shear-band formation where the system becomes increasingly sensitive to failure mechanisms that are beyond the range of our theory.  Thus, we focus primarily on the early onset of ASB and not on the later stages of this phenomenon. 

In Sec.~\ref{DATA1}, we discuss our use of the MD early-stage stress-strain data to determine the material-specific parameters.  Then, in Sec.~\ref{ASB}, we specify the perturbed initial conditions that we use to trigger ASB formation.  We also show our computed spatial distibutions of temperature and plastic deformation during the onset of the  banding instability, and compare these with the experimental measurements where available.  We conclude in Sec.~\ref{CONCLUSIONS} with some remarks about the significance of these calculations.
 
\section{Equations of Motion}
\label{EOM}

A recent, more complete discussion of the following equations of motion can be found in \cite{JSL-arXiv17}.

As in \cite{JSL-17} we consider a strip of polycrystalline material, of width $2W$, driven in simple shear in the $x$ direction at constant velocities $V_x$ and $-V_x$ at its top and bottom edges respectively. The total strain rate is $V_x/W \equiv Q/\tau _0$, where $\tau _0 = 10^{-12}$s is a characteristic microscopic time scale. In order to observe shear localization, we look at spatial variations in the $y$ direction, perpendicular to the $x$ axis.

The local, elastic plus plastic strain rate is $\dot{\varepsilon}(y)=dv_x/dy$, where $v_x$ is the material velocity in the $x$ direction. This motion is driven by a time-dependent, spatially uniform, shear stress $\sigma$. Because the overall shear rate is constant, we can replace the time $t$ by the accumulated total shear strain, say $\varepsilon$, so that $\tau _0 \partial /\partial t \to Q \partial /\partial \varepsilon$. Then we denote the dimensionless, $y$-dependent plastic strain rate by $q(y,\varepsilon )\equiv \tau _0 \dot{\varepsilon }^{pl}(y,\varepsilon )$. The internal state variables that describe this system are the areal density of dislocations $\rho \equiv \tilde{\rho }/b^2$ (where $b$ is the length of the Burgers vector), the effective temperature $\tilde{\chi}$ (in units of a characteristic dislocation energy $e_D$), and the ordinary temperature $\tilde{\theta }$ (in units of the pinning temperature $T_P = e_P /k_B$, where $e_P$ is the pinning energy defined below). Note that $\rho $ also may be interpreted as the total length of dislocation lines per unit volume and $1/\sqrt{\rho }$ is the average distance between dislocations. All three of these dimensionless quantities $\tilde{\rho }$, $\tilde{\chi }$, and $\tilde{\theta }$ are functions of $y$ and $\varepsilon $.

The central, dislocation-specific ingredient of this analysis is the thermally activated depinning formula for the dimensionless plastic strain rate $q$ as a function of a non-negative stress $\sigma$:  
\begin{equation}
\label{qdef}
q(\varepsilon) = \sqrt{\tilde\rho} \,\exp\,\Bigl[-\,\frac{1}{\tilde\theta}\,e^{-\sigma/\sigma_T(\tilde\rho)}\Bigr]. 
\end{equation}
This is an Orowan relation of the form $q = \rho\,b\,v\,\tau_0$ in which the speed of the dislocations $v$ is given by the distance between them multiplied by the rate at which they are depinned from each other.  That rate is approximated here by the activation term in Eq.~(\ref{qdef}), in which the energy barrier $e_P$ (implicit in the scaling of $\tilde\theta$) is reduced by the stress dependent factor $e^{-\sigma/\sigma_T}$, where  $\sigma_T(\tilde\rho)= \mu_T\,\sqrt{\tilde\rho}$ is the Taylor stress, and $\mu_T \equiv r\, \mu$.  The dimensionless number $r$ is the ratio of a depinning length to the length of the Burgers vector.  Thus, $r$ should be approximately independent of temperature and strain rate.  Note that only the magnitude of $\sigma$ appears in this expression for a local time scale.  Directional information would be included in tensorial equations of motion for stress fields and flow patterns, but not in this expression for a scalar time scale.   

The pinning energy $e_P$ is large, of the order of electron volts, so that $\tilde\theta$ is very small.  As a result, $q(\varepsilon)$ is an extremely rapidly varying function of $\sigma$ and $\tilde\theta$.  This strongly nonlinear behavior is the key to understanding yielding transitions and shear banding as well as many other important features of polycrystalline plasticity.  For example, the extremely slow variation of the steady-state stress as a function of strain rate discussed in  \cite{LBL-10} is the converse of the extremely rapid variation of $q$ as a function of $\sigma$ in Eq.~(\ref{qdef}). In what follows, we shall see that this temperature sensitivity of the plastic strain rate is the key to understanding important aspects of the thermomechanical behavior.   

The equation of motion for the scaled dislocation density $\tilde\rho$ describes energy flow. It says that some fraction of the power delivered to the system by external driving is converted into the energy of dislocations, and that that energy is dissipated according to a detailed-balance analysis involving the effective temperature $\tilde\chi$.  This equation is: 
\begin{equation}
\label{rhodot}
\frac{\partial\tilde\rho}{\partial\varepsilon}= \kappa_1\,\frac{\sigma\,q}{\nu(\tilde\theta,\tilde\rho,Q)^2\,\mu_T\,Q}\, \Bigl[1 - \frac{\tilde\rho}{\tilde\rho_{ss}(\tilde\chi)}\Bigr],
\end{equation}
where $\tilde\rho_{ss}(\tilde\chi) = e^{- 1/\tilde\chi}$ is the steady-state value of $\tilde\rho$ at given $\tilde\chi$.  The coefficient $\kappa_1$ is an energy conversion factor that, according to arguments presented in  \cite{LBL-10,JSL-17} and \cite{JSL-arXiv17}, should be approximately independent of both strain rate and temperature for the situations considered here.  The other quantity that appears in the prefactor in Eq.~(\ref{rhodot}) is
\begin{equation}
\label{nudef}
\nu(\tilde\theta,\tilde\rho,Q) \equiv \ln\Bigl(\frac{1}{\tilde\theta} \Bigr) - \ln\Bigl[\ln\Bigl(\frac{\sqrt{\tilde\rho}}{Q}\Bigr)\Bigr].
\end{equation}

The equation of motion for the scaled effective temperature $\tilde\chi$ is a statement of the first law of thermodynamics for the configurational subsystem:  
\begin{equation}
\label{chidot}
\frac{\partial\,\tilde\chi}{\partial\varepsilon} = \kappa_2\,\frac{\sigma\,q}{\mu_T\,Q}\,\Bigl( 1 - \frac{\tilde\chi}{\tilde\chi_0} \Bigr). 
\end{equation}
Here, $\tilde\chi_0$ is the steady-state value of $\tilde\chi$ for strain rates appreciably smaller than inverse atomic relaxation times, i.e. much smaller than $\tau_0^{-1}$. The dimensionless factor $\kappa_2$ is inversely proportional to the effective specific heat $c_{e\!f\!f}$. Unlike $\kappa_1$, there is no  reason to believe that $\kappa_2$ is a rate-independent constant.  In \cite{JSL-17a}, $\kappa_2$ for copper was found to decrease from $17$ to $12$ when the strain rate increased by a factor of $10^6$.  Since we shall consider changes in strain rate of the same order here, we shall assume for simplicity that $\kappa_2$ is a constant.  

The equation of motion for the scaled, ordinary temperature $\tilde\theta$ is the usual thermal diffusion equation with a source term proportional to the input power. We assume that, of the three state variables, only $\tilde{\theta}$ diffuses in the spatial dimension $y$. Thus,   
\begin{equation}
\label{thetadot}
\frac{\partial\tilde\theta}{\partial\varepsilon} = K(\tilde\theta) \,\frac{\sigma\,q}{Q} +\frac{K_1}{Q}\frac{\partial ^2\tilde{\theta }}{\partial y^2}- \frac{K_2}{Q}\,(\tilde\theta - \tilde\theta_0).
\end{equation} 
Here, $K(\tilde\theta) = \beta/ (T_P\,c_p\,\rho_d)$ is a thermal energy conversion factor, while $K_1=k_1 \tau_0/(c_p\, \rho_d)$ characterizes heat conduction in the axial direction of the tube. $c_p$ is the thermal heat capacity per unit mass, $\rho_d$ is the mass density, $0< \beta < 1$ is a dimensionless constant known as the Taylor-Quinney factor, and $k_1$ the thermal conductivity. $K_2$ is a thermal transport coefficient that controls how rapidly the system relaxes toward the ambient temperature $T_0$, that is, $\tilde\theta \to \tilde\theta_0 = T_0/T_P$. As discussed in \cite{Le2017},  $K(\tilde\theta)$ may be non-trivially temperature dependent which, in the range of temperature under consideration, is taken to be
\begin{equation}\label{Kfunction}
K(\tilde\theta) = c_0+c_1\,e^{- c_2/(T_P\,\tilde\theta)}.
\end{equation}
We assume that $K_1$ and $K_2$ are constants, independent of the strain rate and temperature. 

It remains to write an equation of motion for the stress $\sigma (\varepsilon )$ which, to a very good approximation, should be independent of position $y$ for this model of simple shear. Such an equation was derived by one of us in \cite{JSL-17} under the assumption that the shear modulus $\mu $ does not depend on temperature. However, if the temperature rises by 600\,$^\circ$C during ASB formation as reported in \cite{Marchand1988}, then such dependence could be essential. Therefore we start with Hooke's law $\sigma =\mu (\tilde{\theta })[ \varepsilon (y)-\varepsilon^{pl} (y)]$, where, now, $\mu $ depends on the ordinary temperature $\tilde{\theta }$. Differentiating this equation with respect to $\varepsilon $, we get
\begin{equation}
\label{sigmadot}
\frac{\partial \sigma }{\partial \varepsilon}=\mu (\tilde{\theta })[ \frac{\tau _0}{Q}\frac{dv_x}{dy}-\frac{q(y,\varepsilon )}{Q}]+\mu ^\prime (\tilde{\theta }) \frac{\partial \tilde{\theta }}{\partial \varepsilon }[ \varepsilon (y)-\varepsilon^{pl} (y)].
\end{equation} 
Neglecting the second term as small compared with the first and averaging over the width, we obtain
\begin{equation}
\label{sigmadotfinal}
\frac{\partial \sigma }{\partial \varepsilon}=\mu (\tilde{\theta }_0)-\int_{-W}^{+W} \frac{\mu (\tilde{\theta })}{2W}\frac{q(y,\varepsilon )}{Q}\, dy.
\end{equation}
When integrating the first term on the right-hand side of (\ref{sigmadot}) by parts, the term containing the derivative $\mu ^\prime (\tilde{\theta })$ is neglected as being small compared to the remaining terms. Also, we assume Dirichlet boundary conditions: $\tilde{\theta }(\pm W)=\tilde{\theta }_0$. Note that  Eq.~(\ref{sigmadotfinal}) differs slightly from Eq.~(II.11) of \cite{JSL-17} because here the dependence of shear modulus on temperature is taken into account. For the numerical solution of Eqs.~(\ref{thetadot}) and (\ref{sigmadotfinal}) it is convenient to introduce the dimensionless coordinate $\tilde{y}=y/W$. Then, Eqs.~(\ref{thetadot}) and (\ref{sigmadot}) keep their form if $\tilde{y}$ is substituted for $y$, while $K_1 \to \tilde{K}_1=k_1\tau_0/(c_p\,\rho_d W^2)$.

To complete our model of the ASB experiments, we need to specify an instability-triggering mechanism analogous to the long, shallow ``notch'' that Marchand and Duffy inscribed along their equivalent of our $x$ axis. We do this -- somewhat arbitrarily -- by choosing a $\tilde y$ dependent value of the initial effective temperature:  
\begin{equation}
\label{perturbation}
\tilde{\chi }(0,\tilde{y})=\tilde\chi_i - \delta \exp (-\tilde{y}^2/2y_0^2),
\end{equation}
where $\delta $ and $y_0$ are the depth and width of the perturbation. In using this formula we usually have set $\delta, y_0 \ll 1$ to describe a small notch that is about as deep as it is wide.  We emphasize that neither the notch itself nor its assumed dimensions are necessarily realistic aspects of our model and that, for present purposes, we do not need them to be so.   

\section{Data Analysis}
\label{DATA1}

The experimental results of Marchand and Duffy \cite{Marchand1988} for steel HY-100, referred to from here on simply as ``steel,'' along with our theoretical results based on the preceding  equations of motion, are shown by the stress-strain curves in Figs. 1-4.  In each of these figures, the points represent the MD data and the solid curves are our theoretical results including the $\tilde y$-dependent initial perturbation defined in Eq.~(\ref{perturbation}). The dashed curves are the partial fits to the small-strain data that we have used to determine the system parameters as discussed in the following paragraphs.  

In order to compute the theoretical stress-strain curves, we need values for ten system-specific parameters, two initial conditions, and the two perturbative parameters introduced in Eq.~(\ref{perturbation}).  The ten basic parameters are the following: the activation temperature $T_P$, the stress ratio $r$, the steady-state scaled effective temperature $\tilde\chi_0$, the two dimensionless conversion factors $\kappa_1$ and $\kappa_2$, the three coefficients $c_0$, $c_1$, and $c_2$ defining the function $K(\tilde\theta)$ in Eq.~(\ref{Kfunction}), and the two thermal factors $\tilde{K}_1$ and $K_2$. Among them $\tilde{K}_1$ can be determined directly from the known material characteristics of steel and the geometry of the tube.  The thermal diffusivity of steel HY-100 is $k_1/(c_p\,\rho_d) = 9 \times 10^{-6}$\,m$^2$\,s$^{-1}$ \cite{Holmquist1987}, and $W = 1.25$\,mm; thus we find $\tilde{K}_1=5.76\times 10^{-12}$. We also need initial values of the scaled dislocation density $\tilde\rho_i(0,\tilde{y}) \equiv \tilde\rho_i(\tilde{y})$ and the effective temperature $\tilde\chi_i(0,\tilde{y}) \equiv \tilde\chi_i(\tilde{y})$; all of which are determined by sample preparation -- presumably the same for all samples, but possibly a source of experimental uncertainty.  For the ordinary temperature we suppose that $\tilde\theta(0,\tilde y)=\tilde\theta_0$.  Finally, we need a formula for the temperature-dependent shear modulus $\mu(T)$, which we take from \cite{Chen1996} to be  
\begin{equation}
\label{musteel}
\mu(\tilde\theta) = \mu_1 - \left[\frac{D}{\exp(T_1/T_P\,\tilde\theta)-1}\right],
\end{equation}
where $\mu_1 = 7.146 \times 10^4$\,MPa, $D = 2910$\,MPa and $T_1 = 215$\,K. 

\begin{figure}[htp]
	\centering
	\includegraphics[width=.7\textwidth]{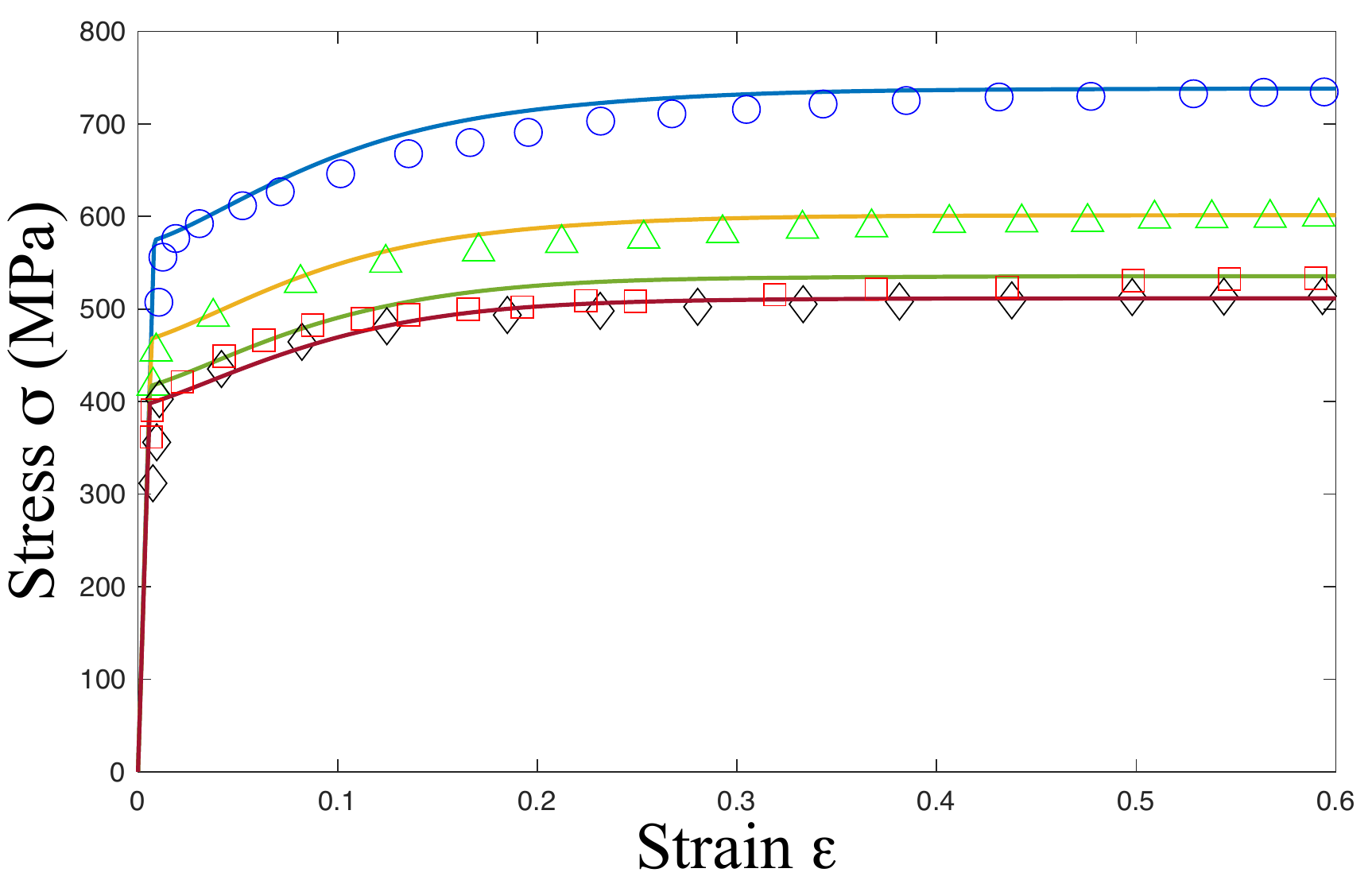}
	\caption{(Color online) Quasi-static stress-strain curves for steel at the strain rate $\dot\varepsilon = 10^{-4}/$s, for temperatures -190\,$^\circ $C, -73\,$^\circ $C, 25\,$^\circ $C, 70\,$^\circ $C shown from top to bottom. The experimental points are taken from Marchand and Duffy \cite{Marchand1988}}
	\label{fig:1}
\end{figure}

In earlier papers starting with \cite{LBL-10}, we were able to begin evaluating the parameters by observing steady-state stresses $\sigma_{ss}$ at just a few strain rates $Q$ and ambient temperatures $T_0 = T_P\,\tilde\theta_0$, and inverting Eq.~(\ref{qdef}) to find 
\begin{equation}
\label{sigmass}
\sigma_{ss} = r\,\mu\,\sqrt{\tilde\rho_{ss}}\,\,\nu(\tilde\theta_0,\tilde\rho_{ss},Q);~~~~\tilde\rho_{ss} = e^{- 1/\tilde\chi_0}.
\end{equation}
Knowing $\sigma_{ss}$, $T_0$ and $Q$ for three stress-strain curves, we could solve this equation for $T_P$, $r$, and $\tilde\chi_0$, and check for consistency by looking at other steady-state situations. With that information, it was relatively easy to evaluate $\kappa_1$ and $\kappa_2$ by directly fitting the full stress-strain curves.  This strategy does not work here because the thermal effects are highly nontrivial.  Examination of the experimental data shown in the figures indicates that all of these samples are undergoing thermal softening at high strain rates and large strains; the stresses are decreasing slowly and the temperatures must be increasing.  Even the curves that appear to have reached some kind of steady state have not, in fact, done so at their nominal ambient temperatures.  

Another possible strategy is to use the version of Eq.(\ref{sigmass}) that is valid just at the yield stress $\sigma_y$, where the deformation is switching abruptly from elastic to plastic:
\begin{equation}
\label{sigmay}
\sigma_{y} = r\,\mu\,\sqrt{\tilde\rho_i}\,\nu(\tilde\theta_0,\tilde\rho_i,Q).
\end{equation}
Again, we could use measurements of $\sigma_y$ to determine $T_P$, $r$, and $\tilde\rho_i$. Here, however, the problem is that, as seen in the Figures at the higher strain rates, these curves exhibit appreciable stress overshoots that make it difficult to evaluate the yield stresses accurately. In fact, our computed curves are presumably consistent with Eq.(\ref{sigmay}), but we have found it best not to rely exclusively on data at the yieldpoints.  

\begin{figure}[htp]
	\centering
	\includegraphics[width=.7\textwidth]{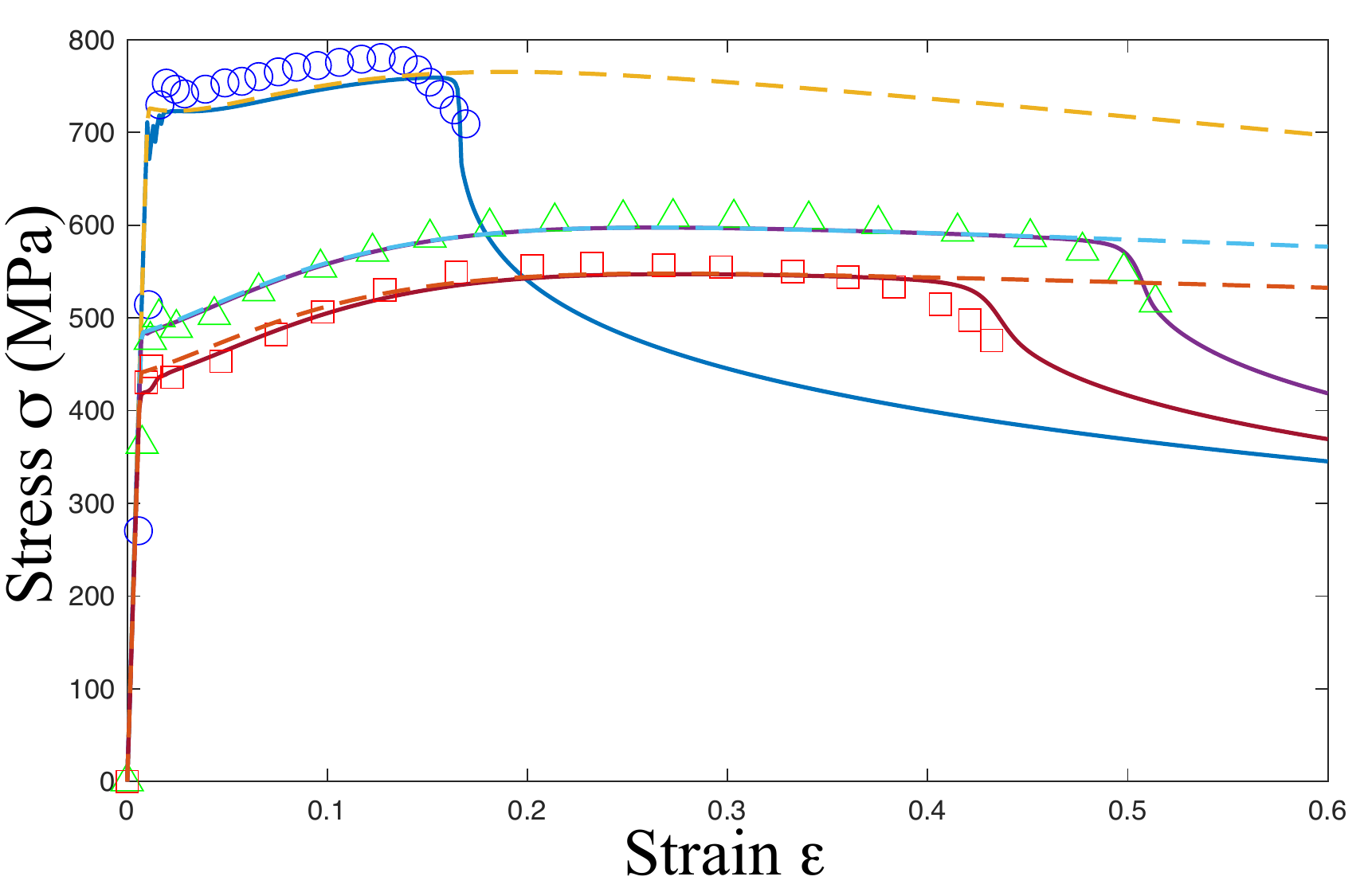}
	\caption{(Color online) Stress-strain curves for steel at the strain rate $\dot\varepsilon =1000/$s, for temperatures -190\,$^\circ $C, 25\,$^\circ $C, 134\,$^\circ $C shown from top to bottom. The experimental points are taken from Marchand and Duffy \cite{Marchand1988}}
	\label{fig:2}
\end{figure}

To counter these difficulties, we have resorted to the large-scale least-squares analyses that we used in \cite{Le2017,LE-TRAN-17}. That is, we have computed the sum of the squares of the differences between our theoretical stress-strain curves and a large set of selected experimental points, and have minimized this sum in the space of the unknown parameters. In order that this procedure be computationally feasible, we have assumed that the observed  stress-strain curves are independent of the notch-like perturbations during the early stages of these experiments, i.e. during what MD call stages I and II in Figure 12 of their paper \cite{Marchand1988}.  We have used only this early-stage data in our fitting procedure.  In this way, we need to solve only the four ordinary differential equations considered in \cite{Le2017}. That is, we set $\delta = 0$ in Eq.~(\ref{perturbation}) and neglect the $\tilde y$ dependence of our state variables.  The results are shown by the dashed curves in our Figures.  

\begin{figure}[htp]
	\centering
	\includegraphics[width=.7\textwidth]{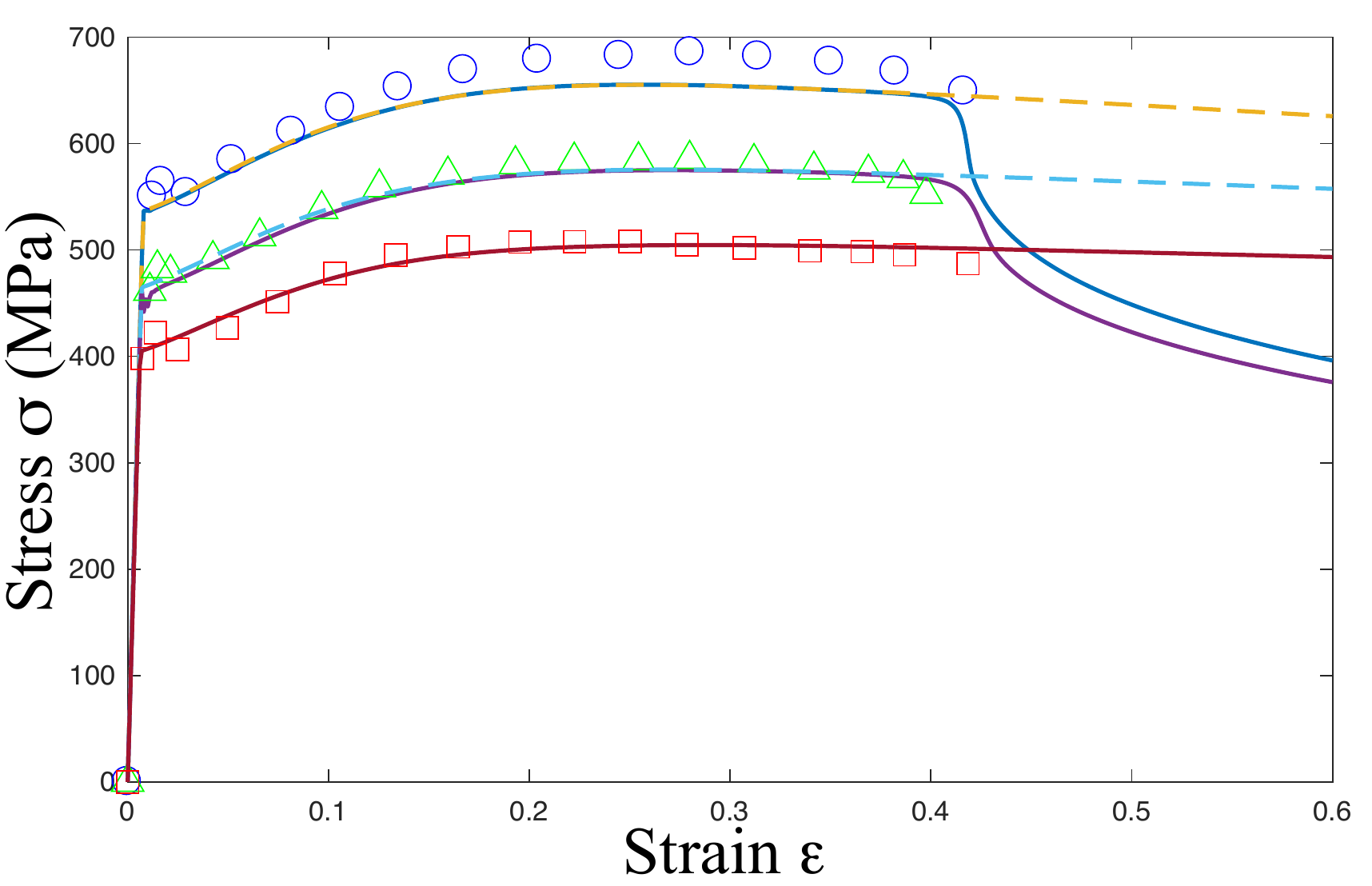}
	\caption{(Color online) Stress-strain curves for steel at the strain rate $\dot\varepsilon =1000/$s , for temperatures -73\,$^\circ $C, 70\,$^\circ $C, 250\,$^\circ $C shown from top to bottom.  The experimental points are taken from Marchand and Duffy \cite{Marchand1988}}
	\label{fig:3}
\end{figure}

With just one exception, we find that all twelve of the MD early-stage stress-strain curves can be fit with just a single set of system parameters.  These are: $T_P = 5.16\times 10^5$\,K,\,$r = 0.0178,\,\chi_0= 0.229,\,\kappa_1=7.65,\,\kappa_2=14.3,\,c_1 = 4 \times 10^{-7}$(MPa)$^{-1}$,~$c_2= 2\times 10^{-7}$(MPa)$^{-1}$,\,$c_3 = 400$\,K,\,$\tilde\rho_i=0.0076$, $\tilde\chi_i=0.212$, and $K_2=1.66 \times 10^{-14}$.  The single exception is that, at the lowest temperature reported by MD, $-\,190^\circ$C, and for the only strain rate reported at that temperature, $10^3$\,s$^{-1}$, we find that we need a somewhat larger value of the initial dislocation density, $\tilde\rho_i=0.0097$ instead of $0.0076$ in order to fit the data.  It seems to us that this agreement between theory and experiment is well within the bounds of experimental uncertainty.  

Here is one useful check on the internal consistency of this analysis. With $\rho _d = 7748$\,kg/m$^3$  and $c_p = 502$\,J/kg\,K (see \cite{Holmquist1987}), we find that our function $K(\tilde\theta)$ implies a maximum Taylor-Quinney factor $\beta \approx 0.99$ that is slightly smaller than unity within the range of temperatures under consideration.  

\section{Adiabatic Shear Banding}
\label{ASB}

\begin{figure}[htp]
	\centering
	\includegraphics[width=.7\textwidth]{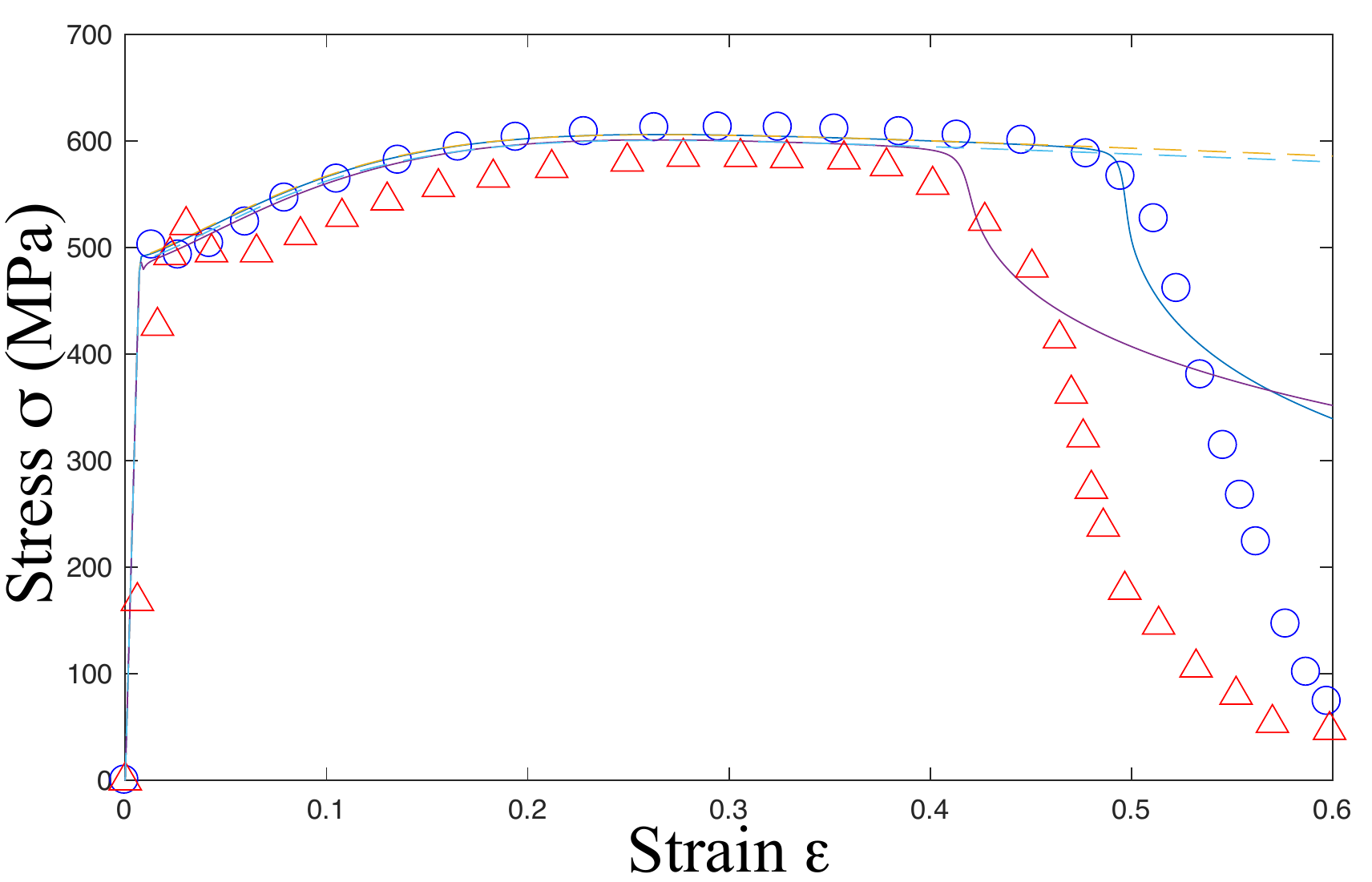}
	\caption{(Color online) Stress-strain curves for steel at the strain rates $\dot\varepsilon =3300/$s and $\dot\varepsilon =1600/$s shown from top to bottom, for room temperature. The experimental points are taken from Marchand and Duffy \cite{Marchand1988}}
	\label{fig:4}
\end{figure}

\begin{figure}[htp]
	\centering
	\includegraphics[width=.7\textwidth]{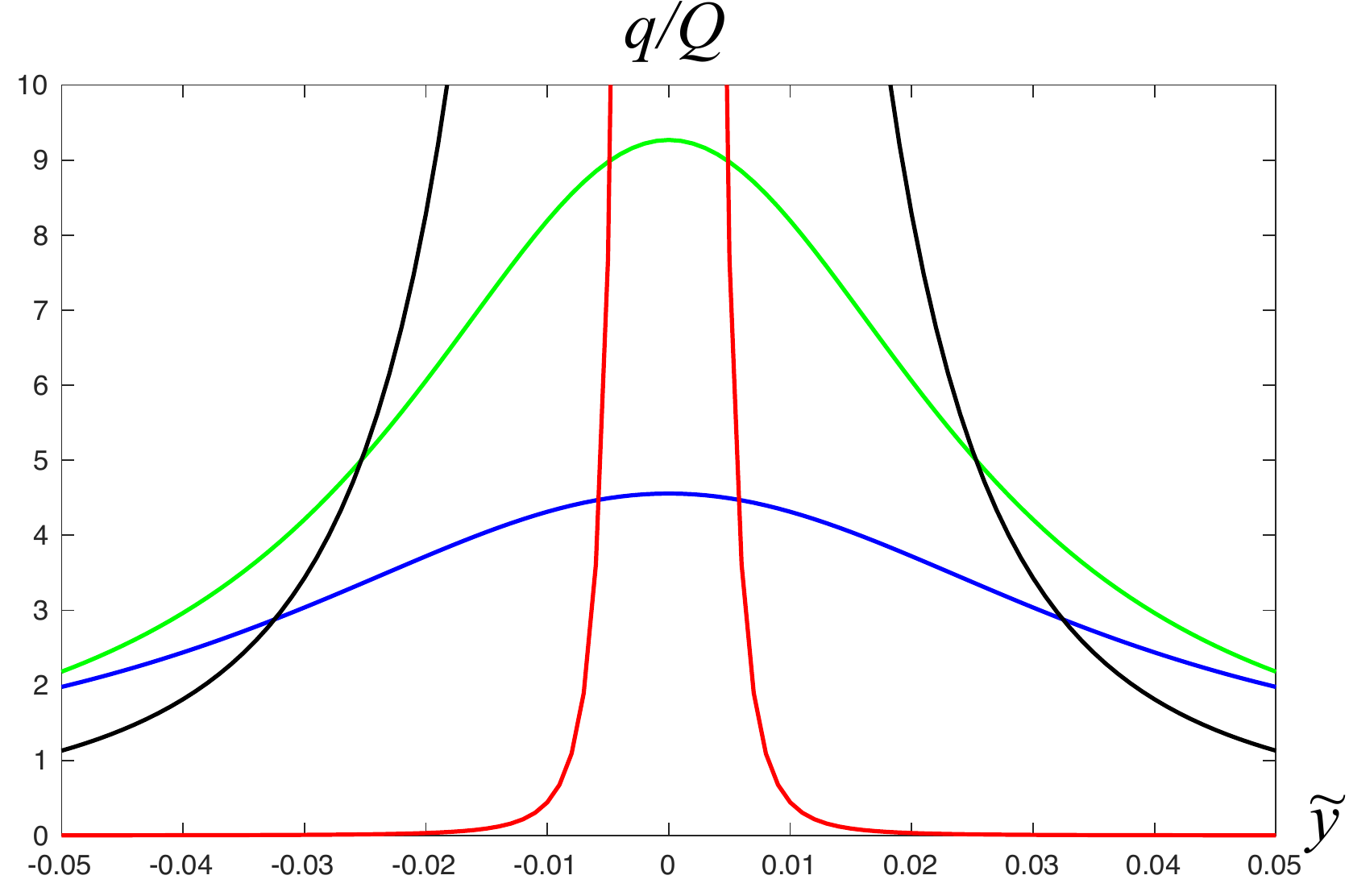}
	\caption{(Color online) Theoretical fractional strain rate distributions for steel for $\dot\varepsilon =3300/$s at room temperature: (i) at strain $\varepsilon =0.45$ (blue), (ii) at strain $\varepsilon =0.47$ (green), (iii) at strain $\varepsilon =0.49$ (black), (iv) at strain $\varepsilon =0.497$ (red)}
	\label{fig:5}
\end{figure}

\begin{figure}[htp]
	\centering
	\includegraphics[width=.7\textwidth]{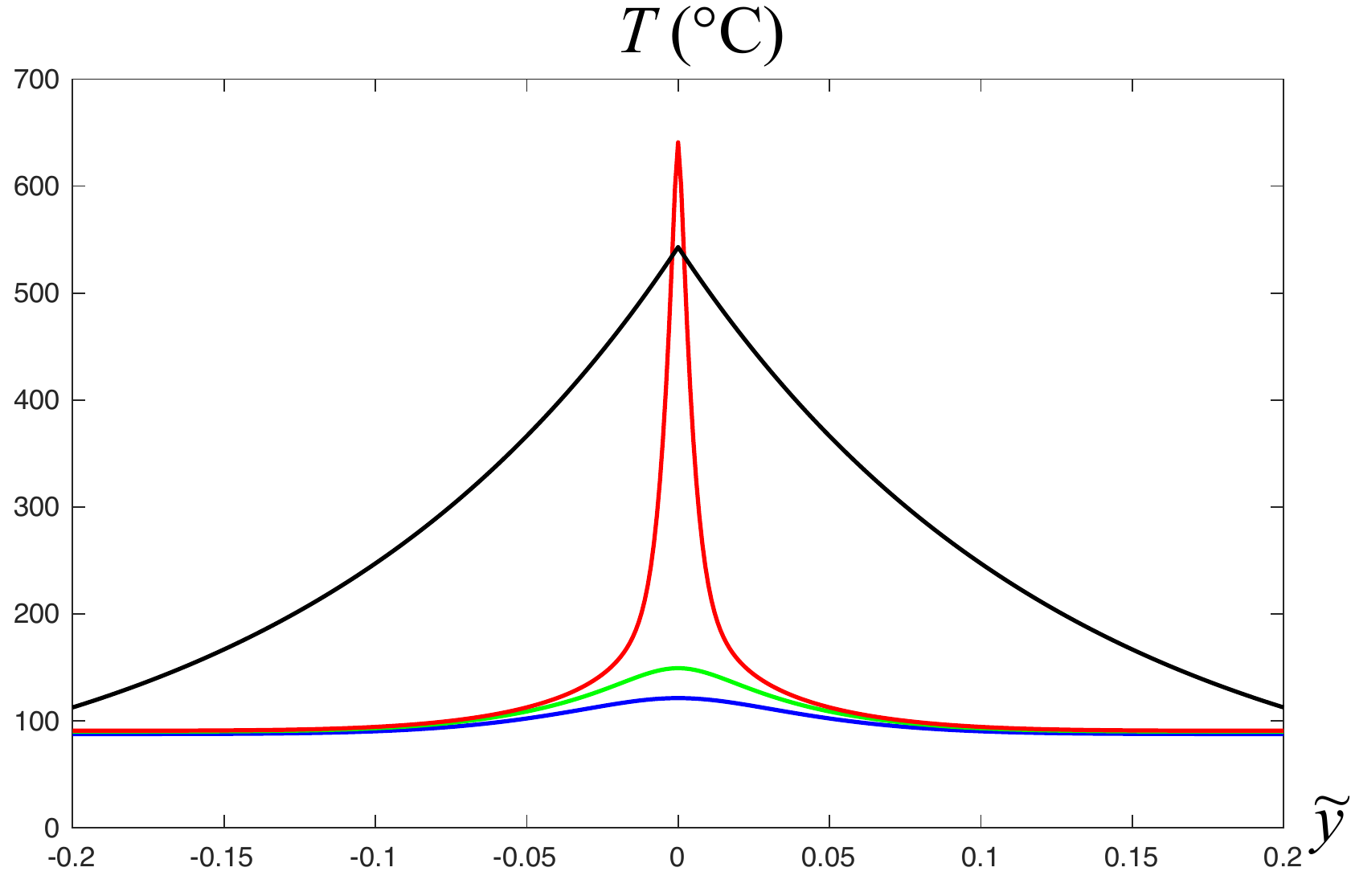}
	\caption{(Color online) Temperature distributions for the strain rate $\dot\varepsilon =3300$/s and room temperature at the strains $\varepsilon =0.46$ (blue), $\varepsilon =0.48$ (green), and $\varepsilon =0.505$ (red). The black curve is the empirical law proposed by Marchand and Duffy \cite{Marchand1988}}
	\label{fig:6}
\end{figure}

\begin{figure}[htp]
	\centering
	\includegraphics[width=.7\textwidth]{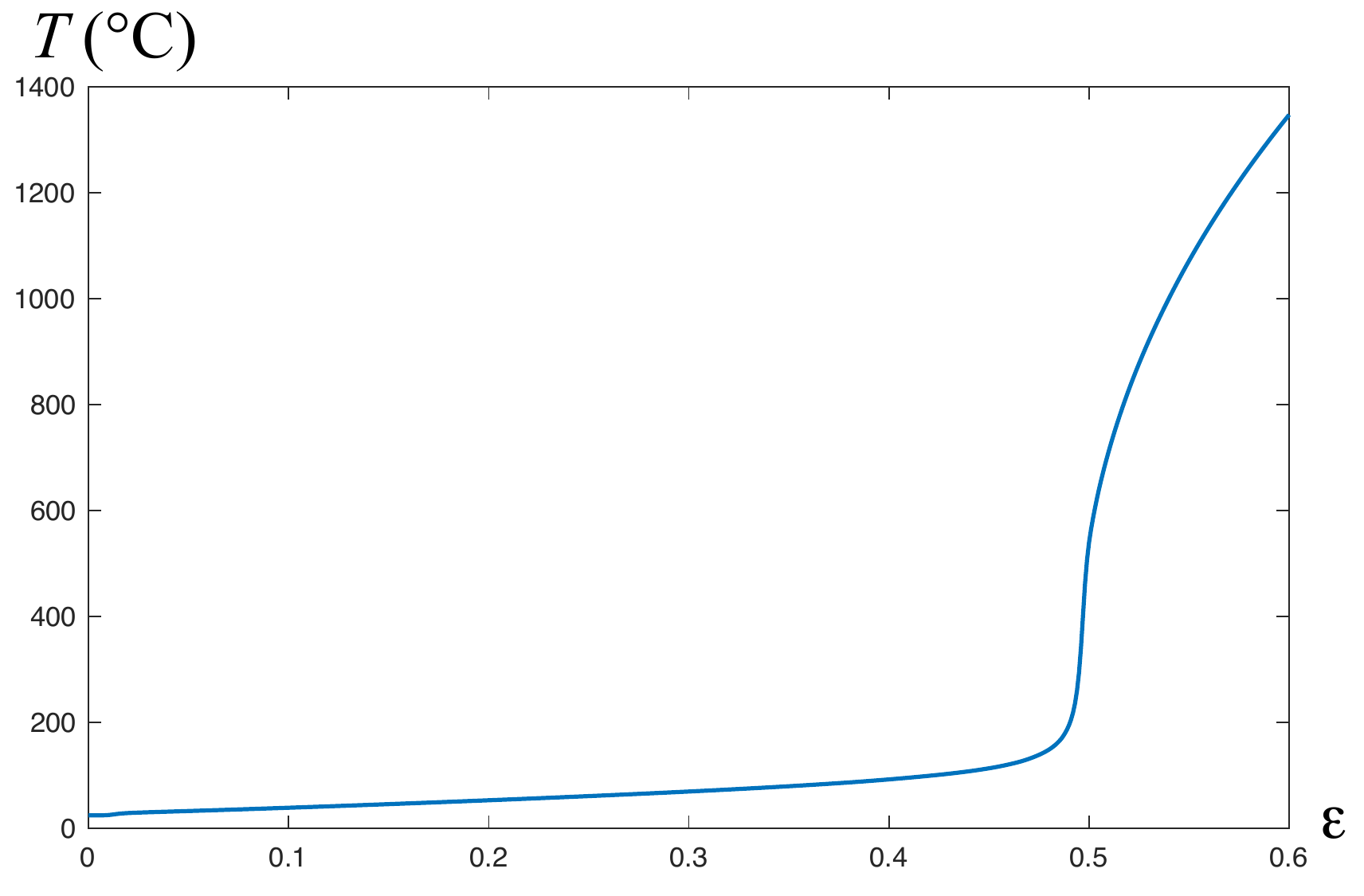}
	\caption{(Color online) Temperature at the center of the shear band ($\tilde{y}=0$) for the strain rate $\dot\varepsilon =3300$/s and room temperature as the function of the strain.}
	\label{fig:7}
\end{figure}

We turn now to the main topic of this paper -- the onset and early development of adiabatic shear banding instabilities.  To study these phenomena theoretically, we introduce into our equations of motion the $\tilde y$-dependent notchlike initial perturbation defined in Eq.~(\ref{perturbation}).  Now, our dynamical variables $\tilde\rho$, $\tilde\chi$, and $\tilde\theta$ become functions of $\tilde y$, and we need to solve the complete system of integro-differential equations (\ref{rhodot})-(\ref{sigmadotfinal}) subject to initial and boundary conditions. To do this, we discretize the equations in the interval $(-1< \tilde y <1)$ by dividing it into $2n$ sub-intervals of equal length $\Delta \tilde{y}=1/n$. Then the second spatial derivative of $\tilde{\theta }$ in equation (\ref{thetadot}) is approximated by
\begin{equation}
\frac{\partial ^2\tilde{\theta }}{\partial \tilde{y}^2}(\tilde{y}_i)=\frac{\tilde{\theta }_{i+1}-2\tilde{\theta }_i+\tilde{\theta }_{i-1}}{(\Delta \tilde{y})^2},
\end{equation}
where $\tilde{\theta }_i=\tilde{\theta }(\tilde{y}_i)$. Similarly, the integral over $\tilde{y}$ of any function $f(\tilde{y})$ is computed by using the trapezoidal rule
\begin{equation}
\int_{-1}^{1} f(\tilde{y})\, d\tilde{y}=\Delta \tilde{y} [f_{-n}/2+f_{-n+1}+\ldots +f_{n-1}+f_n/2] .
\end{equation}
In this way,  we reduce the four integro-differential equations to a system of $6n+1$ ordinary differential equations. We have solved these numerically using the Matlab-ode15s solver with $n = 1000$ and the $\varepsilon$ step equal to $0.001$. Those solutions are shown by the solid lines in Figs.~\ref{fig:1}-\ref{fig:4} and by the graphs of strain rate and temperature as functions of $\tilde y$ in Figs.~\ref{fig:5}-\ref{fig:7}.  

\begin{table}
  \centering 
\begin{tabular}{|c|c|c|c|c|c|c|c|c|}
\hline
$T$\,($^\circ$C) & -190 & -73 & 25 & 25 & 25 & 70 &  134 & 250 \\
$\dot{\varepsilon }$\,(1/s) & 1000 & 1000 & 1000 & 1600 & 3300 & 1000 & 1000 & 1000\\  
\hline
$\delta $ & 0.08 & 0.026 & 0.0271 & 0.0396 & 0.0172 & 0.059 & 0.0823 & - \\ \hline
$y_0$ & 0.0362 & 0.026 & 0.03 & 0.042 & 0.0295 & 0.059 & 0.0705 & - \\ \hline
\end{tabular}  
  \caption{The values of $\delta$ and $y_0$}\label{table:1}
\end{table}

As shown in these figures, the perturbation produced by the notch does not affect the overall stress-strain relation either for very small, quasistatic strain rates or during the early stages of the more rapid shear deformations.  This is what we assumed when arguing in favor of our parameter-fitting procedure.  In the quasistatic cases, any extra heat generated near the notch diffuses away quickly on the time scale of the inverse shear rate.  During the early stages of the faster deformations, it takes appreciable times (in strain units) before the nonlinear instabilities near the notch grow enough to destabilize the system as a whole.  That time before onset, or equivalently the strain at which the stress begins to drop abruptly, is strongly sensitive to the strain rate, the temperature, and the strength of the notch.  The values of $\delta$ and $y_0$ that we have chosen to fit the observed onsets are shown in Table 1.  We emphasize again that the irregularity of these values is almost certainly an experimental artifact.  There is no reason to believe that our notch model is realistic enough or that Marchand and Duffy could control their initial conditions accurately enough to expect greater precision.

The highly nonlinear onset of banding is seen most clearly in the graphs of the computed fractional strain rate $q(\varepsilon,\tilde y)/Q$ shown in Fig.~\ref{fig:5} for $\dot\varepsilon = 3300$\,s$^{-1}$ at room temperature, $T \cong 25\,^\circ$C. The apparent onset strain is $\varepsilon \cong 0.50$.  However, at $\varepsilon = 0.45$, $q/Q$ at $\tilde y = 0$ has increased by only a factor of about 2, and the width of the perturbation has not increased appreciably from its initial value of $2\,y_0 \cong 0.1$.  By $\varepsilon = 0.47$, $q/Q$ has grown by another factor of $2$ but the width has not changed, nor has $q/Q$ changed from its initial value of unity outside the emerging band.  Finally, by $\varepsilon = 0.49$, the band has begun to sharpen dramatically.  By $\varepsilon = 0.497$, it has collapsed into a narrow region of width approximately $0.01$ at $\tilde y \cong 0$. $q/Q$ has vanished outside the narrow band, and the total strain rate $Q$ is now concentrated inside the band with the maximum of $q/Q$ achieving 1000 in the middle of the band (not shown on the Figure). The overall, uniform stress has dropped to the value that is needed to drive the highly concentrated plastic flow at a high temperature.

Several graphs of temperature $T=T_P\,\tilde\theta(\varepsilon,\tilde y)-273$\,K (in Celsius) for $\dot\varepsilon = 3300$\,s$^{-1}$ at the ambient room temperature, $T \cong 25\,^\circ$C, are shown in Fig.~\ref{fig:6}, and a graph of the temperature at the center of the band as a function of $\varepsilon$ is shown in Fig.~\ref{fig:7}.  By $\varepsilon \cong 0.505$ which is the onset of ASB,  the latter temperature has reached about 600\,$^\circ$C, which is consistent with the value reported by MD as shown in their Fig. 20. This consistency is significant.  It is based on independently determined thermal parameters; thus it is a sensitive test of the strong nonlinearity of the theory. We also show in Fig.~\ref{fig:6} the empirical law proposed by MD, $T=a\, e^{-7.875 |\tilde{y}|}$ ($a=543\,^\circ$C), that is based on several measurements of temperature at somewhat different strain rates. This empirical distribution of temperature is wider than our simulated distribution, possibly because it was measured at a significantly larger strain where more heat had been generated in the band and had diffused into the neighboring material.

The experimental stress drops following the onset of banding are generally much deeper than predicted by our theory.  This seems reasonable because, as mentioned in the Introduction, there must be other physical mechanisms  that come into play in this regime.  There is probably a transition between shear banding and fracture, during which the resistive  mechanism switches from high-temperature plasticity to friction between the faces of two separate materials in contact with each other.   

\section{Concluding Remarks}
\label{CONCLUSIONS}

The main conclusion of this paper is that the statistical thermodynamic dislocation theory provides an accurate picture of adiabatic shear banding as observed by Marchand and Duffy \cite{Marchand1988}.  The theory's description of the coupling between stress, temperature and strain rate accounts quantitatively for the strong thermal instabilities leading to abrupt stress drops seen in those experiments. 

More generally, we are addressing two different kinds of issues here, one pertaining to first-principles theoretical physics and the other to applied materials research.  On the one hand, we are testing the validity of the thermodynamic dislocation theory.  On the other, we are trying to find ways in which these new insights can be used to predict the performance of materials in engineering applications. 

Because our theoretical starting point is unconventional, we have made special efforts to construct and test it as rigorously as possible.  We have imposed stringent requirements on our equations of motion and on our choice of the parameters that appear in them.  The equations themselves are statements of well known physical principles -- conservation of energy and flow of entropy in accord with the second law of thermodynamics -- and they are expressed in terms of properly defined internal state variables -- the dislocation density and the two thermodynamically defined temperatures.  We postulate no phenomenological fitting functions.  Each of the parameters that occur in our equations can, in principle, be determined either by independent measurement or first-principles computation.  

We have started with a strong additional postulate designed to test the validity of this theory. Specifically, we have assumed that almost all of our fundamental parameters remain constant across the wide range of strain rates and temperatures that we are exploring.  This postulate includes conversion factors like $\kappa_1$ and $\kappa_2$, which we know to be variable in some circumstances, and it also includes initial conditions that are subject to uncertainties of sample preparation.  Nevertheless, this postulate works remarkably well. The twelve stress-strain curves shown in Figs. 1-4 are all in reasonable agreement with experiment.  The main exception is the top curve in Fig. 2 where we have slightly adjusted the initial value of the dislocation density $\tilde\rho_i$ as discussed at the end of Sec.~\ref{DATA1}. A similar adjustment of one of the $\tilde\rho_i$'s in Fig.~4 might have improved the agreement with experiment for the two curves shown there.  The important point, however, is that these minor disagreements are much more likely to have been caused by experimental inaccuracies than by systematic errors in the theory.  

Similarly, the stress overshoots at the initial yield points that appear in all the high-strain-rate experimental MD curves in Figs. 2-4 are almost certainly instrumental effects having to do with the sudden onset of shear stress.  As shown in Fig.~7 of \cite{JSL-arXiv17}, we can reproduce those overshoots simply by reducing slightly the initial values of the scaled effective temperature $\tilde\chi_i$.  But doing so would require readjusting the $\tilde\chi_i$'s for the quasistationary cases where no overshoots occur. Thus we conclude that, like the failure during late-stage ASB, the initial stress overshoots are caused by physical mechanisms outside the range of our theory. 

In our opinion, by far the most important outstanding questions are those pertaining to physical interpretations of our theory and, thus, to connections between the theory and its applications.  The skeptical reader will have noticed that terms such as ``cross slip,'' ``stacking fault,'' ``dislocation pile-up,'' or even ``crystal symmetry'' or ``glide plane'' do not appear in this paper.  We argue that physical concepts like these belong in first-principles calculations of dynamic quantities such as $\kappa_1$, $\kappa_2$ or the parameter $r = \mu_T/\mu$ that determines the relation between stress and plastic strain rate. The standard practice in the conventional literature has been to try to go directly from observed microscopic behaviors of small groups of dislocations to phenomenological models of strain hardening or fracture toughness.  That strategy, however, has not succeeded in producing usefully predictive theories. The present line of  investigation seems to us to be more promising.  

\medskip
\noindent Acknowledgments.

\medskip
T.M. Tran acknowledges support from the Vietnamese Government Scholarship Program 911.  JSL was supported in part by the U.S. Department of Energy, Office of Basic Energy Sciences, Materials Science and Engineering Division, DE-AC05-00OR-22725, through a subcontract from Oak Ridge National Laboratory.

\end{document}